# Biomechanical comparison between manual and motorless device assisted patient handling: sitting to and from standing position

Jean-Baptiste RICCOBONI [a, b, c], Tony MONNET [b], Antoine EON [b], Patrick LACOUTURE [b], Jean-Pierre GAZEAU [b], Mario CAMPONE [c]

[a] *Institut de Biomécanique Humaine Georges Charpak, 151 Boulevard de l'Hôpital, Paris, France*
[b] *Institut PPRIME, 11 Marie et Pierre Curie boulevard, Futuroscope Chasseneuil, France*
[c] *Institut de Cancérologie de l'Ouest, 15 Rue André Boquel, Angers, France*

*Keywords:* Biomechanical, Caregiver, Constraint, Device, Disorders, Handling, Health, Load, Medical, Musculoskeletal, Nurse, Patient, Transferring.

## ABSTRACT

*Background:* Occupational safety and health institutions report that caregivers are particularly at risk of developing work-related musculoskeletal disorders (WRMSDs) and patient handling is often pointed out as one of the main causes. While lots of studies addressed the use of assistive devices in order to protect caregivers, it seems that motorless stand-up lifts have not been studied yet.

*Objectives:* The aim of this work is to provide quantitative data about the loads in the low back area, as well as qualitative data about subjects perceptions, resulting from the use of a motorless stand-up lift and to compare them to those resulting from manual patient handling.

*Methods:* Nine caregivers participated to motion capture and ground reaction forces measurement sessions. These recordings were performed in three cases of handling: manual handling with one caregiver, manual handling with two caregivers, motorless device assisted handling. Forces and torques at the L5/S1 joint were then estimated through Inverse Dynamics process. A questionnaire about manual and motorless device assisted handling was also submitted.

*Results:* Motorless device assisted handling involved the smallest loads whereas manual handling with one caregiver involved the biggest loads.

*Conclusions:* Our findings suggest that, if the situation allows it, caregivers should be helped by another caregiver or use a motorless stand-up lift when handling a patient from sitting to standing position or from standing to sitting position considering the reduced loads these aids involve.

# 1. Introduction

It is now a well-known fact that people working in the healthcare sector, involved in some way in patient handling (called caregivers in the article), are particularly at risk of developing work-related musculoskeletal disorders (WRMSDs), especially at the back, and therefore to be impacted by all their usual consequences (Bureau of Labour Statistics, 2016; Health and Safety Executive, 2016/17 (1); Health and Safety Executive, 2016/17 (2)). Examples can be found in the works of Applegeet et al. (2007), Hignett et al. (2016) and Ribeiro et al. (2017).

As pointed out by Roffey et al. (2010), it is still difficult to show that workplace manual handling or assisting patients are independently causative of pain at specific areas. WRMSDs are complex problems, with many and various parameters. A few instances of these, while not directly linked to patient handling, are discussed in the articles of Carneiro et al. (2017), Holman et al. (2010), Kay et al. (2015), Manyisa et al. (2017) and Risør et al. (2017).

Nevertheless, detecting and identifying the elements of patient handling leading to too much stress (compared to the acceptable values recommended by occupational safety and health institutions) then modifying them in order to decrease, nullify or avoid their related constraints (with the expectation of reducing the people exposure to WRMSDs) has been the subject of lots of studies these last years.

About detecting and identifying, questionnaires (Mirmohammadi et al., 2015), physical (Hwang et al., 2017; Lee et al., 2017; Schall et al., 2016; Yan et al., 2017) or physiological measurements (Wang et al., 2017) and risk assessment methods including several relevant criteria (Villarroya et al., 2016) are often used.

Other studies tried to determine some of the components of WRMSDs appearance: McDermott et al. (2012) on training in patient handling, Lee et al. on availability and use of patient lifting equipment (2013) or more broadly on the importance of organizational safety culture and practices (2017).

Some solutions are then proposed: the most popular consists of training the personnel (Hodder et al., 2010; Resnick et al., 2009), using assistive devices (Alamgir et al.; 2009) or even approaches combining multiple interventions (Haglund et al., 2010; Huffman et al.; 2014; Krill et al., 2012) even though these should be furthered (Thomas et al., 2014). In a similar way, Waters et al. (2007) described high risk tasks and proposed solutions for each.

Finally there are studies that try to compare these solutions and it is in this category that our work falls, focusing on the potential benefits that mechanical assistive devices can get to the problem of WRMSDs. Even if other technologies to measure biomechanical constraints are promising, like inertial measurement units (Bolink et al., 2016; Faber et al., 2016; Koopman et al., 2018), electrogoniometers (Weiner et al., 2017) and computer vision (Mehrizi et al., 2018), most of the studies in this category have used optoelectronic motion capture system. Dutta et al. (2012), Garg et al. (1991), Santaguida et al. (2005) and Zhuang et al. (1999) have worked on a couple of devices going from walking belts to overhead lifts and compared the biomechanical constraints involved in each case. Zhuang et al. (2000) have also done a psychophysical assessment of some assistive devices. Yet and to our knowledge, it seems that one type of device have not been studied: the motorless stand-up lifts.

The aim of this work is to provide quantitative data about the loads in the low back area, as well as qualitative data about subjects perceptions, resulting from the use of a motorless stand-up lift and to compare them to those resulting from manual patient handling.

## 2. Methods

### a. Subjects

#### i. Caregivers

A group of nine caregivers, composed of seven women and two men, volunteered and were therefore selected after giving their informed consent in accordance with local ethical procedures. They were from 21 to 59 years old (mean: 38.3; standard deviation: 14.2), were from 155 to 181 centimeters tall (169.7; 9), weighed from 48 to 98 kilograms (66.1; 14.7). There were five nurses and four assistant nurses; all of them knew how to manually handle patient (at least from their initial training courses) and how to use the motorless stand-up lift; all of them were working at the Institut de Cancérologie de l'Ouest (ICO), a institute specialized in cancer treatments (both outpatient and inpatient), research and training. They had from 0.7 to 41 years of experience (17.2; 15). Seven of them suffered or had suffered from MSDs, mainly at the low back and shoulders areas. They claimed that they manually handle patients 9.5 times a day on average (with a standard deviation of 6.3) and that this require 1.6 persons on average (0.5). They used the motorless stand-up lift 0.8 times a day (1.7) and it requires 1.3 persons (0.5).

#### ii. Surrogate patient

Each caregiver played the role of the handled patient (called the surrogate patient in this article) several times, meaning they had to imitate the behaviour of real patients. Their instructions were to provide the minimum efforts allowing the handling to be carried out. In addition and for the purpose of providing a point of comparison between the subjects, a male surrogate patient (called MSP) measuring 175 centimeters and weighing 63 kilograms participated in all the experiments with the same instructions. More detailed explanations will be given afterwards.

### b. Setting and equipment

The experiments took place in a motion capture room at the Institut PPRIME. As these measures were not *in situ*, we used a seat with an adjustable (and not tracked) height in order to mimic an hospital bed in terms of functionality and clutter. This height changed between subjects.

The motorless stand-up lift used was a Vertic'Easy from HMS-VILGO (Fig. 1).

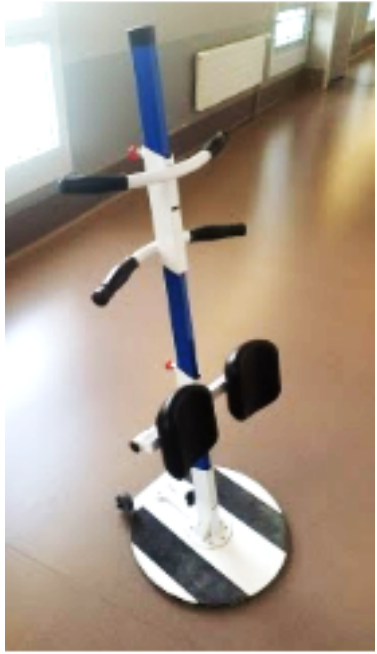

**Fig. 1.** Motorless stand-up lift

Its dimensions was 1300mm in height, 510mm in length, 470mm in width and it weighed 16kg unloaded. The position of the handgrips and tibial support were not tracked (as determining the best parameters to handle patient was not an objective of this study) but did not changed during and between experiments, even if the subjects were told that they were allowed to change them depending on what they thought was most comfortable.

    c.   Measurement techniques

By measuring the three-dimensional location of body segments and the mechanical forces applied by the subject on the ground then by using Inverse Dynamics process, it is possible to calculate the joint reaction forces and torques. We also wanted to obtain several details about the subjects so we would notably be able to compare our results to those of previous studies.

    i.   Motion capture system

A Vicon Motion Systems Limited (Oxford, GB) motion capture system composed of 19 MX T40 cameras sampled at 100Hz, connected through Vicon Motion Systems Limited's Nexus 1.8.5, was used for all the recordings.

Subjects were equipped with 52 reflective markers following a modified version of the Istituto Ortopedico Rizzoli Full-Body Marker Set (Cappozzo et al., 1995; Leardini et al., 2011), showcased in Fig. 2. Head and upper limbs segments had their movements recorded with a view to describe full-body postures in future studies but were not used in this one nor included in the biomechanical constraints computation.

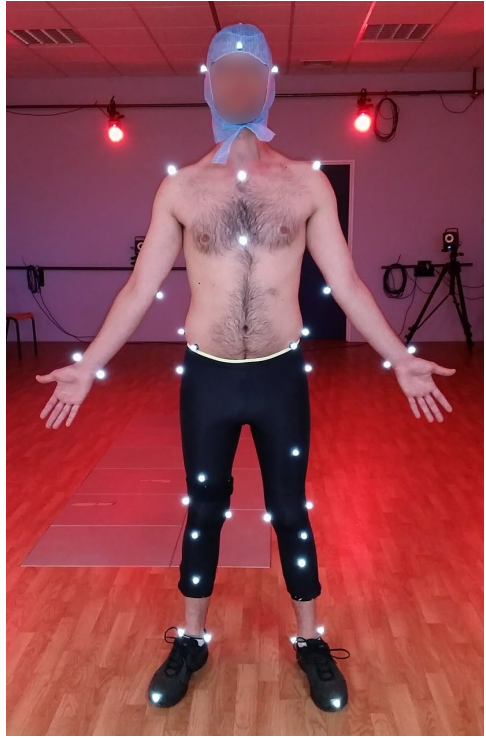
**Fig. 2.** Subject equipped with markers

    ii.    Ground reaction forces measuring instrument

Ten M40-600x400 force platforms from SENSIX (Poitiers, FR) sampled at 1000Hz were placed at the center of the recording and calibrated area and were synchronized with the kinematics acquisitions by Vicon Motion Systems Limited's Nexus 1.8.5. Their layout allowed subjects to put their feet "naturally", in their own words, while having one foot per platform during handling.

    iii.    Subject data procurement technique

A questionnaire was submitted to the subjects at the end of each session. Despite the fact that not all the questions and their answers will be described in this article, here are the main topics covered: background information (like gender, age, position, seniority…), presence or not of WRMSDs and details about it if so, quantity and perception of manual and motorless device assisted patient handlings. The objective of this questionnaire was to strengthen the analysis of the biomechanical results.

    d.  Handling methods

Prior to the experiments, we spent one month at the ICO premises to get a better understanding of all the tasks performed by caregivers. The following handling methods were chosen from the observations made at the ICO and from the advices of the first participating caregivers.

While the techniques employed by the subjects were highly similar, their gestures presented some variations regarding hands positions and orientations in particular. We decided to let them some freedom on this in order to obtain the most natural behaviour.

i. Manual handling with one caregiver (MH1C)

Subjects were asked to lift alone the surrogate patients by facing them and placing their legs on each side of those of the surrogate patients. The Fig. 4a and 4b illustrate this handling where the subject chose to place his hands on the back of the surrogate patient with his arms under the armpits. In Fig. 4c, the subject chose to place his hands lower on the back, one above the other.

ii. Manual handling with two caregivers (MH2C)

When two caregivers were involved, the one equipped with the markers (called the main caregiver in this article) was on the left of the surrogate patient (Fig. 4d and 4e) while the other was on the right. Both were free on how to lift the surrogate patient as long as the feet of the main caregiver were entirely on two different force platforms. The right arm of the main caregiver was placed under the left armpit of the surrogate patient and the right hand was on the left forearm.

iii. Motorless device assisted handling (MDAH)

The subject faced the surrogate patient when using the motorless stand-up lift. The latter was in the middle, the tibial support on the side of the surrogate patient. The subject immobilized the device thanks to the upper handgrips (Fig. 4f) while the surrogate patient used the lower handgrips as a support to stand up. Pictures of a MDAH are shown in Fig. 3.

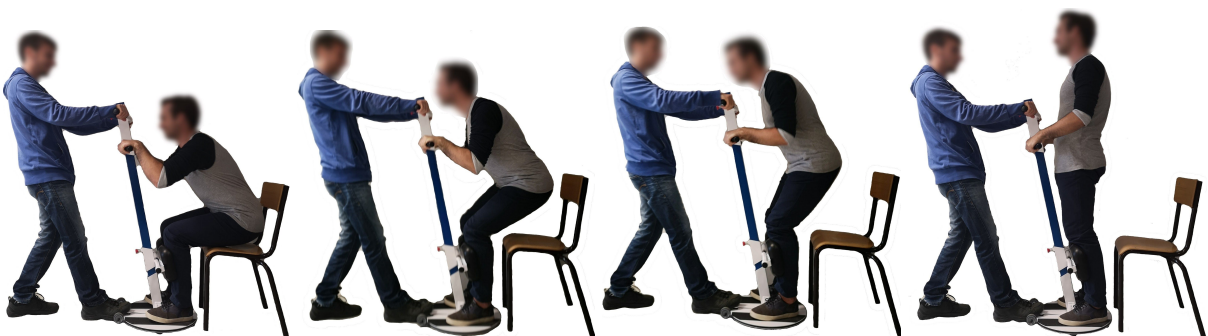

**Fig. 3.** Pictures of a MDAH

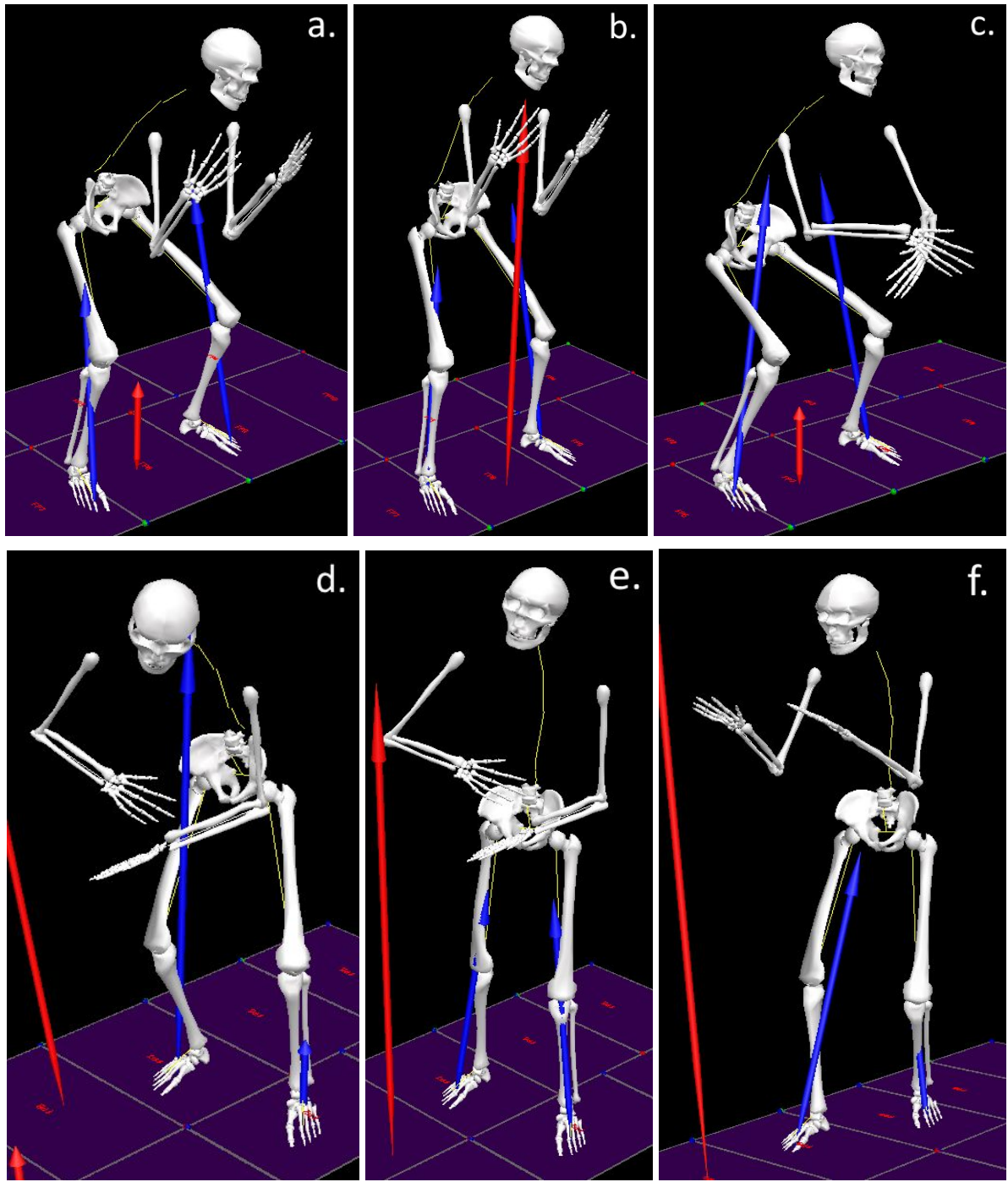

**Fig. 4.** Examples of handlings obtained through Visual3D, from top left to bottom right: start of MH1C (a.), end of it (b.), start of MH1C showing a different position of the hands (c.), start of MH2C (d.), end of it (e.), MDAH (f.)

e. Data collection procedure

Caregivers arrived by group of three at the Institut PPRIME (all recordings took place over three days). They had been previously informed of the general conduct of the recording session, were asked to come with tight clothes and gave their consent through a signed agreement. At their arrival, we presented them more thoroughly the different steps of the day. Subjects were equipped with the markers one at a time and we proceeded every time to a few trials of manual and motorless device assisted handling before starting to record. Once the subjects felt ready, we started with manual handlings. The equipped subject (playing the role of the main caregiver) was asked to lift the two others subjects and the male surrogate patient (MSP) up alone three times each (one by one). The main caregiver was next helped by a second caregiver to perform the manual handlings of the last remaining subject and the MSP; then the second caregiver and the remaining subject switched roles. Finally, the main caregiver handled the two other subjects and the MSP with the motorless stand-up lift. Surrogate patients were sitting on a seat at a height set by the main caregiver. Each subject performed 3 trials for the 10 handling cases. An example of this description is in the Table 1. We removed afterwards the markers from the main caregiver and positioned them on one of the two other subjects (who became the main caregiver) to restart all the procedures described above. When all the recordings for the three subjects were done, we gave them the questionnaire to fill as their last task.

| Handling method | Caregiver(s) | Surrogate patient | | |
|---|---|---|---|---|
| | | S2 | S3 | MSP |
| MH1C | S1 | 3 trials | 3 trials | 3 trials |
| MH2C | S1 and S2 | / | 3 trials | 3 trials |
| | S1 and S3 | 3 trials | / | 3 trials |
| MDAH | S1 | 3 trials | 3 trials | 3 trials |

**Table 1.** Example of trials recording (S1, S2, S3 are subjects; S1 is the only one equipped with markers; MSP is the male surrogate patient)

f.  Data analyses

i.  Biomechanical measurement

Kinematics and ground reaction forces data were exported as a .c3d file then imported into C-Motion's (Germantown, MD) Visual3D. Beforehand, we had built a custom kinetic model in which segments and joints are considered as rigid bodies and spherical joints respectively. Joints positions were defined as follow: midpoints of the malleoli for the ankles, midpoints of the femoral epicondyles for the knees, results of regression equations for hips adapted from Bell et al. (1989; 1990); the back have been divided into five segments with the six joints being from bottom to top: the fifth, third and first lumbar vertebrae, the seventh and third thoracic vertebrae and the seventh cervical vertebra. Segments definition of volumes, mass, moments of inertia and center of gravity location were inspired by the work of Ernest P. Hanavan (1964). Associating feet to their corresponding force platforms and performing the Inverse Dynamics process, we obtained joint forces and torques for each segment in its frame of reference. These results were finally compared in the three cases (MH1C, MH2C and MDAH). To improve data readability, each trial was divided in two parts: when subjects helped the surrogate patient to stand up (called the standing up part) and when subjects helped the surrogate patient to sit (called the sitting part).

ii. Answers to the questionnaire

Answers to the questionnaire were used to know, inter alia, subject's perceptions and were compared to biomechanical measurement.

iii. Statistical analysis

One-way ANOVAs were run for forces and torques to compare standing up and sitting parts and to compare handling methods. The significance level was 0.05.
Multiple comparisons were then performed (Tukey's test) to test for differences among pairs of means.

## 3. Results

a. Loads

Fig. 5 and Fig. 6 (for sitting and standing part respectively) present forces and torques at the L5/S1 joint of one of the subjects using each handling method to handle the MSP (weight: 63kg) each time. Graphs for other subjects had similar looks. Abscissa axis is in percentage of movement completion (0 beginning and 100 end of the movement) and ordinate axis are in N or N.m.

For the sitting part, maximum values for forces and torques were close between both manual handling methods. These values however were smaller with the motorless handling method. Forces remained constant overall through handling for the three methods while torques increased through handling for the manual handlings and were nearly constant for the MDAH. When looking at the motion files, the raise in forces and torques from 0 to 5% of movement completion happened when the surrogate patient knee's started to bend. The drops in forces and torques near 80% of movement completion happened when the surrogate patient started to lie on the seat.

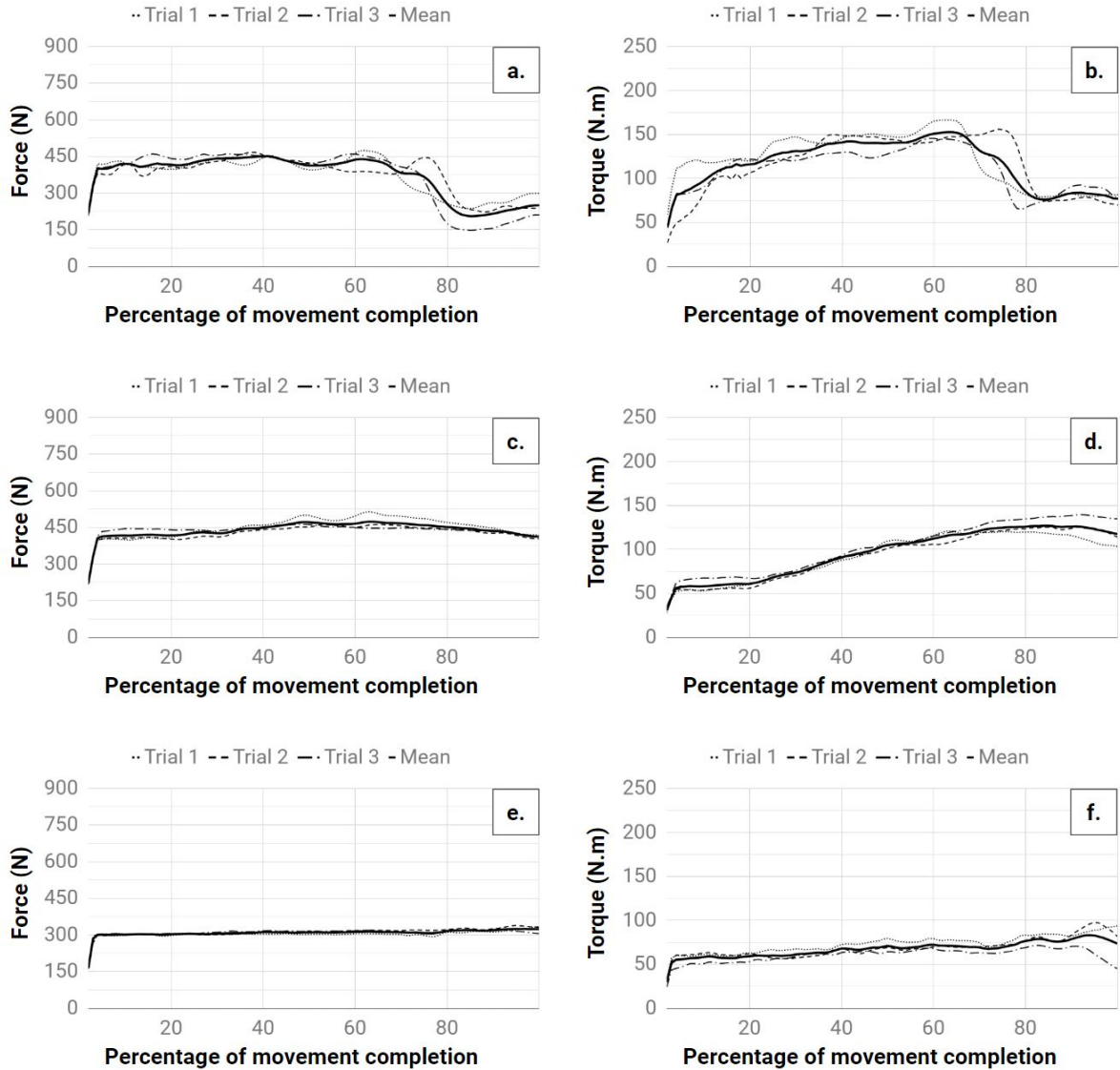

**Fig. 5.** Forces and torques at the L5/S1 joint during the sitting part with:
MH1C (a. and b.), MH2C (c. and d.), MDAH (e. and f.)

For the standing up part, differences between handling methods in force and torque values were bigger. MH1C had the highest maximum values while MDAH had the lowest. These peak values happened around 50% of movement completion for manual handling methods. Like for the sitting part, MDAH induced nearly constant forces and torques. When looking at the motion files, the raise in forces and torques from 0 to 5% of movement completion happened when the main caregiver started to pull the surrogate patient up.

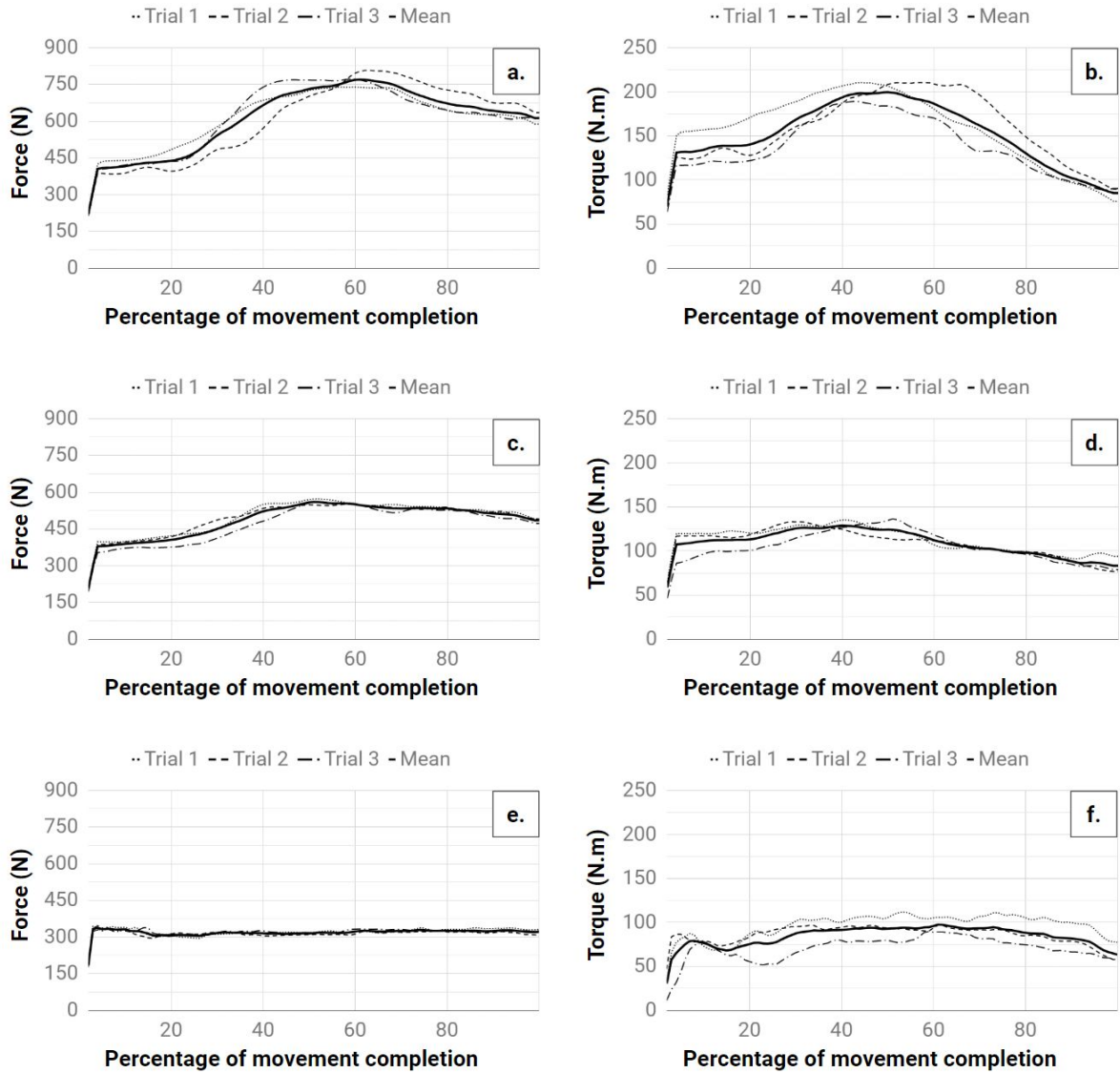

**Fig. 6.** Forces and torques at the L5/S1 joint during the standing up part with: MH1C (a. and b.), MH2C (c. and d.), MDAH (e. and f.)

Fig. 7 and Fig. 8 illustrate the comparisons between the sitting and standing parts and between handling methods respectively. Bars represent the means of the maximum values of forces and torques for all the corresponding trials.

When comparing sitting and standing up parts, forces and torques were significantly bigger during the standing up part with the manual handling methods and there were no significant differences with the MDAH method. Mean forces increased from 471.1N to 613.8N (F=170.9; p<0.05) with one caregiver, from 454.8N to 537.2N (F=130.6; p<0.05) with two caregivers and were the same (353.3N≈350.2N; F=1.53; p=0.22) with one caregiver and the motorless stand-up lift. Mean torques increased from 114.8N.m to 148.2N.m (F=74.53; p<0.05) with one caregiver, from 94.6N.m to 103.0N.m (F=8.69; p<0.05) with two caregivers and were the same (64.4N.m≈68.8N.m; F=1.26; p=0.26) with one caregiver and the motorless stand-up lift.

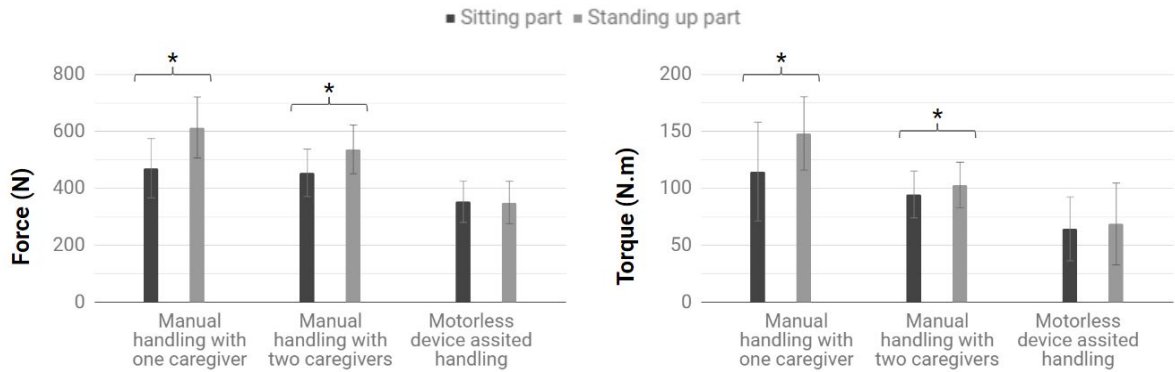

**Fig. 7.** Comparison between sitting and standing up part with respect to forces and torques at the L5/S1 joint for each handling method

For the sitting part: changing handling method induced significant differences for forces (F=186.7; p<0.05) and torques (F=118.5; p<0.05). Multiple comparisons revealed that forces differed with a p-value of 0.04 for the MH1C/MH2C comparison and with a p-value <0.01 for the MH1C/MDAH and MH2C/MDAH comparisons. It revealed that torques differed with a p-value of <0.01 for the three comparisons. For the standing up part: changing handling method induced significant differences for forces (F=493.4; p<0.05) and torques (F=217.1; p<0.05). Multiple comparisons revealed that forces and torques differed with p-values <0.05 for the three comparisons.

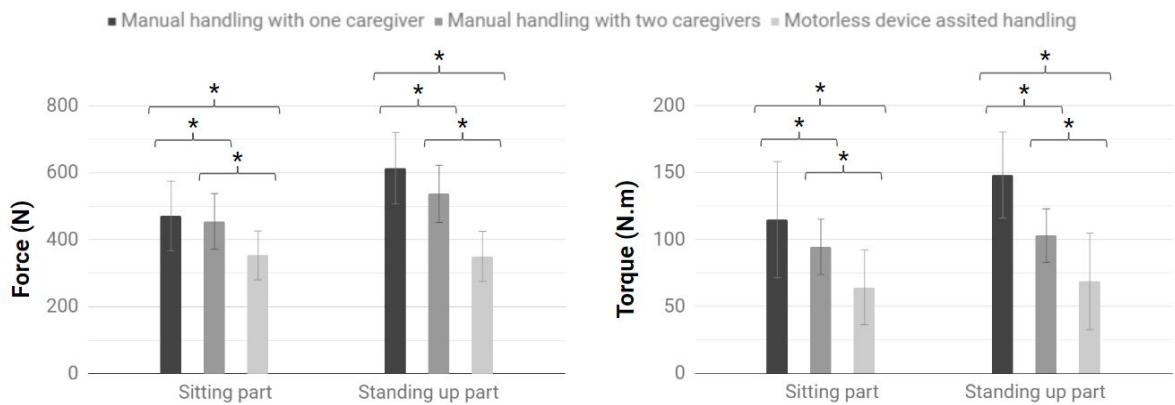

**Fig. 8.** Comparison between handling methods with respect to forces and torques at the L5/S1 joint for both trial parts

b. Questionnaire

Table 2 provides a tally of the caregivers' answers to the questionnaire.

| Question | Answer |
| --- | --- |
| Do you suffer or have you ever suffered from musculoskeletal disorders? | Yes currently: 4<br>Yes but not anymore: 3<br>No: 2 |
| If yes, which body are/were concerned? | Shoulder: 4<br>Lower back: 4<br>Upper back: 3 |
| Rate the pain intensity on a scale from 1 (bearable) to 10 (unbearable). | Mean: 4.7<br>Standard deviation: 1.8 |
| Rate on a scale from 1 (demanding) to 5 (not demanding) how you consider manual handling from the caregiver point of view ? | Mean: 3.3<br>Standard deviation: 0.9 |
| Rate on a scale from 1 (not comfortable) to 5 (comfortable) how you consider manual handling from the surrogate patient point of view ? | Mean: 3.3<br>Standard deviation: 0.7 |
| Have you ever used the motorless stand-up lift alone before today ? | Yes: 7<br>No: 2 |
| Rate on a scale from 1 (demanding) to 5 (not demanding) how you consider motorless device assisted handling from the caregiver point of view ? | Mean: 4.3<br>Standard deviation: 0.7 |
| Rate on a scale from 1 (not comfortable) to 5 (comfortable) how you consider motorless device assisted handling from the surrogate patient point of view ? | Mean: 3.7<br>Standard deviation: 0.9 |
| Which type of handling do you prefer as a caregiver? | Motorless device assisted: 5<br>Manual: 4 |
| Which type of handling do you prefer as a surrogate patient? | Motorless device assisted: 5<br>Manual: 4 |

**Table 2.** Answers to the questionnaire

Subjects slightly prefered MDAH to manual handling from the point of view of both the caregiver and the surrogate patient. As we expected tight results, we asked them to explain their answers. In favour of the motorless stand-up lift, their arguments were: easier to use for the personnel, easier to perform lifts without pain and fatigue, reassuring when the patient body weight is important or when the patient physical abilities are diminished, better protection of caregiver health, better control of patient equilibrium. In favour of manual handling, their arguments were: more pleasant for the personnel and less intimidating for the patient, quicker and more adaptable to the environment, causing less cluttering.

## 4. Discussion

### a. MH1C versus MH2C

In light of these results and if the situation allows it, we recommend that caregivers handle patients with the help of one colleague instead of doing it alone. While the load difference is the smallest of our comparisons between handling methods, caregivers may feel a benefit over the long term because reducing the loads have a positive effect on the exposure to WRMSDs by reducing fatigue phenomenon and risk of injury. Loads were also applied less time with MH2C, as shown for instance in Fig. 5b and 5d.

### b. Manual methods versus MDAH

Furthermore, we strongly incite caregivers to use the motorless stand-up lift considering the involved loads. This device helped to greatly reduce forces and torques on the low back area for all the subjects. Using it is even more interesting if there is no other caregiver available to help handling patients as this comparison presents the biggest difference.

### c. Questionnaire

We expected these results because of what we retained from the month spent at the ICO (unpublished data). The low utilization of the motorless stand-up lift by caregivers, compared to the frequency of manual handlings as claimed in section 2.a.i, seemed to be more related to the difficulties in getting it ready for use (e. g. due to its relative unavailability) than to the difficulties of its use. This study could bring more information about its benefits and we felt that caregivers' opinion on handling equipments evolved through our work as they seemed more inclined to use them in the future.

### d. Comparison with previous findings

As written in the introduction, it seems that there are no studies with an aim close to ours. We would like to remind on another note that our computations were done using the ascent method of the Inverse Dynamics process. The values presented here are composed of the effect of the subject upper body (trunk, head, upper limbs) mass and the load applied by the surrogate patient which complicate comparisons with studies using other measurement or computation methods.

That being said, our findings are going in most studies' direction: dividing the loads between caregivers and/or using assistive device help reducing loads and thus fighting against WRMSDs.

e. Limitations

   i. Subjects

To begin with, the number of subjects was quite low. This was mainly due to the fact that it was not possible for the ICO to send more people as it would have been difficult to ensure quality care without enough employees. Means presented in this study may vary with more subjects. Having both genders further diminish the reliability of our comparisons. The subject data (like age, size, weight, experience…) present sometimes important standard deviation and, in combination with the small pool of subject, this could affect our results in unknown ways. Experience with an assistive device, for instance, can help the caregivers to lower the loads on their low backs (Dutta et al.; 2011).

As more than 60% of the caregivers working at the ICO present MSDs (unpublished results), we decided to not exclude them from our selection of subjects because it would have been even more difficult to find participants and also not representative of this population. Subjects suffering or having suffered from MSDs may have behaved differently from others, which may have altered the values of the biomechanical measurements.

It should be noted that all the subjects were able to pull themselves up with the motorless stand-up lift. There are cases where real patients can not do so because they are too weakened, using a motorless stand-up lift is then not possible. Caregivers are often well aware of this fact but it is worth to mention it.

   ii. Setting and equipment

The experiments were done in a motion capture room. Despite our efforts to mimic an hospital room and the numerous trials performed before recording, environment and technical requirements may have impacted subject behaviour.

We tested only one model of motorless stand-up lift. Results may be different with other models and/or brands.

   iii. Data collection

Even if we tried to not exhaust our subjects, our data collection procedure had to be efficient in order to allow the caregivers to get back to their job as soon as possible. Therefore, fatigue may have induce an increasing bias in the values of the biomechanical measurements through the day.

   iv. Handling methods

There are two important points here.

First, we only test one method to handle the surrogate patient for each case of handling, each method described as the best one for its case according to the caregivers. It could be interesting to test several methods for each case and to compare them.

Second, while most of the aspects of each methods were imposed, some (like hands position) were decided by the subject. We believe that the results have not been significantly affected but further research should be done.

v.   Measurement techniques

Our subjects worn shoes during the experiments because they do at the hospital. As described by Debbi et al. (2012), force platforms give an approximate location of the center of pressure and therefore induce an approximation in calculations. Despite having ten force platforms and that the caregivers told they felt no constraints about the placement of their feet, instrumented force shoes could have been interesting to use. Some applications and information about this technology can be found in the work of Faber et al. (2010; 2018).

As pointed out by Ghezelbash et al. (2018) and since our biomechanical model includes posteriorly shifted joints and frictionless spherical joints, our computed spinal loads are likely overestimated. This phenomenon is even more complex when considering that there is a displacement of the center of rotation during motion (Senteler et al.; 2018). Research done by Nerot et al. (2018) could help but is yet limited to the standing position. This should not however change our findings because we worked on comparisons.

Finally, it should be noted that there are side effects when using handling equipment: some risks are discussed in the study of Elnitsky et al. (2014) and Menzel et al. (2004) talk about the problem of loads displacement to wrists and hands. Unfortunately, we did not have the means nor the time to consider this last point for the motorless device assisted handlings. Oakman et al. (2018) tried to widen the WRMSDs issue by highlighting the gaps of modern workplace risk management practices.

## 5. Conclusions

To be helped by an other caregiver or by the use of a motorless stand-up lift when handling a patient from sitting to standing position or from standing to sitting position reduce loads in the low back area. Although sometimes difficult to obtain or apply because of several adverse conditions, these aids are well received by the caregivers. We suggest, if the situation allows it, to use them considering the reduced loads they involve.

## Acknowledgements

This project was conducted as part of a partnership between the Institut PPRIME and the Institut de Cancérologie de l'Ouest (ICO). It was funded by the ICO. We would like to thank all the persons involved in the project, with a particular attention for the caregivers, who were approachable and cooperative right from the start until the very end. Findings, conclusions, views and recommendations expressed in this article do not necessarily reflect those of the ICO. Mention of product names does not constitute endorsement by the Institut PPRIME and the ICO.
This research was partially funded by the French government research program "Investissements d'Avenir" through the Robotex Equipment of Excellence: ANR-10-EQPX-44.